\documentclass[11pt,preprint2]{emulateapj}

\shorttitle{The molecular envelope of VY~CMa}
\shortauthors{Muller et al. 2006}

\begin{document}

\title{The molecular envelope around the red supergiant VY~CMa}

\author{Muller S., 
Dinh-V-Trung, 
Lim J., 
Hirano N.} 
\affil{Academia Sinica Institute of Astronomy and Astrophysics (ASIAA), P.O. Box 23-141, Taipei, 106 Taiwan}
\email{muller, trung, jlim, hirano@asiaa.sinica.edu.tw}
\author{Muthu C.} 
\affil{Aryabhatta Research Institute of Observational Sciences (ARIES), Manora Peak, Nainital 263129, India}
\email{muthu@aries.ernet.in}
\and
\author{Kwok S.} 
\affil{Department of Physics, Faculty of Science, University of Hong Kong}
\email{sunkwok@hkucc.hku.hk}

\begin{abstract}
We present millimeter interferometric observations of the molecular envelope around the red
supergiant \object{VY~CMa} with the SubMillimeter Array (SMA)
\footnote{The Submillimeter Array is a joint project between the Smithsonian Astrophysical Observatory
and the Academia Sinica Institute of Astronomy and Astrophysics and is funded by the Smithsonian
Institution and the Academia Sinica.}. The high angular resolution
($<$ 2$\arcsec$) allows us to derive the structure of the envelope as observed in the 1.3 mm
continuum, $^{12}$CO(2-1), $^{13}$CO(2-1) and SO(6$_5$-5$_4$) lines emission. The circumstellar
envelope is resolved into three components: a dense, compact and dusty central component,
embedded in a more diffuse and extended envelope plus a high velocity component. We construct
a simple model, consisting of a spherically symmetric slowly expanding envelope and bipolar
outflows with a wide opening angle ($\sim$ 120$\degr$) viewed close to the line of sight
(i = 15$\degr$). Our model can explain the main features of the SMA data and previous
single-dish CO multi-line observations. An episode of enhanced mass loss along the bipolar
direction is inferred from our modelling. The SMA data provide a better understanding of the
complicated morphology seen in the optical/IR high resolution observations.

\end{abstract}


\section{Introduction}

The terminal stage in the evolution of massive stars (M $>$ 10 M$_\odot$) is characterized by
very short timescales ($\sim$ 10$^4$ yr) and drastic changes in their immediate
circumstellar environment. In the red supergiant phase, periods of large mass loss release
a thick circumstellar envelope of molecular gas and dust. Understanding the mass loss process
from red supergiants is extremely important in that it will eventually govern their evolution
as well as the structure of their surrounding envelope. In a galactic context, supergiants
contribute to the enrichment of the interstellar medium in dust and heavy elements and participate
to the global galactic chemical evolution.

VY Canis Majoris (VY~CMa) is one of the brightest known red supergiants. Its distance has long
been debated \citep{her70, lad78}, but there is now a consensus on a value of 1.5 kpc,
consistent with the velocity and proper motions of H$_2$O masers \citep{ric98}. The stellar
luminosity is estimated to $\sim$ 2 10$^5$ L$_\odot$ \citep{mon99}. The stellar
mass is thought to be $\ge$ 15 M$_\odot$. The basic properties of VY~CMa are summarized in Table 1.

VY~CMa is characterized by an apparent very high mass loss rate, of order 2 -- 4 10$^{-4}$
M$_\odot$/yr (see {\em e.g.} \citealt{bow83, dan94, sta95, har01}), which produced a very
thick circumstellar envelope of dust and molecules, which almost completely obscurs the
central star. The optical/infrared morpholgy of the nebulosity around VY~CMa is very complex,
showing arcs, filaments, bright knots in a very complicated arrangement \citep{smi01, mon99}.
Long slit spectroscopy \citep{smi04, hum05} reveals a very complex kinematics with evidence
for localized and episodic mass loss events. \citet{hum05} showed that the prominent arcs
are expanding relative to the star and represent discrete ejection events. The ejection dates
are ranging from 200 to 1000 years ago and state that VY~CMa has had a period of major activity
during the past 1000 years. These works however rely on scattered lights which may reflect only
the illuminated part of the nebula and the origin of the features seen in the optical and near
infrared images is very uncertain.

We note that in a recent paper, \citet{dec06} propose a theoretical model, based on a spherically
symmetric envelope, that requires time varying mass-loss rates to reproduce previous single-dish CO
multi-line spectra. The high angular resolution view offered by the SubMillimeter Array (SMA) now
provides new constraints on the structure and kinematics of the circumstellar molecular envelope
of this peculiar object. In this paper, we present our SMA observations and discuss about the
morphology and kinematics of the molecular envelope around VY~CMa. We propose a simple model to
explain our data and compare with observations at other wavelengths.

\section{Observations}

We obtained 1.3 mm interferometric observations of VY~CMa with the SubMillimeter Array (SMA,
\citealt{ho04}), which consists of eight 6 m antennas, located on top of the Mauna Kea, Hawaii.
The phase reference of the array was set to
R.A. = 07$^{\rm h}$22$^{\rm m}$58$\fs$328 and Dec. = $-$25$\degr$46$\arcmin$03$\farcs$19.

VY~CMa was observed at 2 different epochs and with 2 different configurations. The first
observation was carried out on 2004 July 31$^{\rm st}$, with 7 antennas in a compact configuration,
including projected baselines from 8 to 99 m. The system temperatures at 1 mm were between 100
and 200 K for the different antennas. In order to improve the angular resolution, a second
track was observed on 2005 January 19$^{\rm th}$, with 7 antennas in an extended configuration.
The projected baselines ranged from 18 to 192 m. The system temperatures were between 200 and
400 K. The bandpass was calibrated using Uranus and/or Jupiter, while Callisto set the flux
calibration. The nearby quasars 0607$-$157 and 0730$-$116 were regularly observed for the gain
calibration. After calibration, the visibilities from both tracks have been combined together.
The data reduction and calibration were done under MIR/IDL. Imaging and data analysis were
entirely processed under GILDAS\footnote{http://www.iram.fr/IRAMFR/GILDAS/}. 

The heterodyne receivers were tuned to allow multi-line observations including the
$^{12}$CO(2-1) line in the upper side band (USB) and the $^{13}$CO(2-1) and SO(6$_5$-5$_4$)
lines in the lower side band (LSB). The SMA correlator bandwidth is 2 GHz wide for each
sidebands, which are separated by 10 GHz. The instrumental spectral resolution was 0.8125
MHz, but we smoothed it to 3.25 MHz ({\em i.e.} $\sim$ 4 km~s$^{-1}$) for the $^{12}$CO(2-1)
and SO(6$_5$-5$_4$) lines and to 8.125 MHz ({\em i.e.} $\sim$ 11 km~s$^{-1}$) in the case of
$^{13}$CO(2-1).

We obtained the map of the 1.3 mm (225 GHz) continuum emission by averaging the line-free
channels from both LSB and USB, resulting in a total bandwidth of 3 GHz. To maximize the
angular resolution, we applied uniform weighting to the continuum visibilities, leading
to a synthesized beam size of 1.38$\arcsec$ x 1.23$\arcsec$ (PA = 169$\degr$). We estimate
a rms noise level of 8 mJy/beam in this continuum map.

The channel maps of the $^{12}$CO(2-1) and SO(6$_5$-5$_4$) line emission were obtained
using natural weighting and gaussian tapering with a full width at half maximum (FWHM) of
130 m, making a compromise between brigthness sensitivity and angular resolution. The
$^{13}$CO(2-1) map was derived using a taper distance of 80 m to increase the signal-to-noise
ratio. The resulting synthesized beam sizes and 1$\sigma$ noise levels are reported in Table 2.
The continuum visibilities were subtracted before imaging. All the maps were deconvolved
with a standard CLEAN/CLARK algorithm.

At 230 GHz, the FWHM of the 6 m SMA antennas primary beam is 54$\arcsec$. The shortest
projected baseline being 8 m, we may loose informations on structures larger than 40$\arcsec$.
Nevertheless, the resulting coverage of the uv-plane is rather scarce until 18 m. We may
therefore also miss part of the emission associated with structures larger than 18$\arcsec$,
although we show in \S\ref{molecularspectra} from the comparison of the $^{12}$CO(2-1) 
SMA and single dish data, that the missing flux is limited to less than 20 \%.

\section{Results}

\subsection{1.3 mm continuum emission} \label{cont}

\citet{lip05} observed VY~CMa at centimeter wavelengths using the VLA. They do not resolve
the source at an angular resolution of $\sim$ 1$\arcsec$ at 1.3 cm. They also present Keck
mid-infrared imaging and show that the dust emission is elongated in the south-west direction
(PA = 227\degr) with FWHM $\sim$ 0.4$\arcsec$ and 0.3$\arcsec$ at 11.7 and 17.9 $\micron$
respectively. From diffraction-limited speckle interferometry, \citet{wit98} measure a dust
shell size of 138 x 205 mas (2.17 $\micron$), 80 x 116 mas (1.28 $\micron$) and 67 x 83 mas
(0.8 $\micron$) with roughly a north-south elongation. Thus, at near-IR wavelengths, the size
of the dust-emitting region decreases towards shorter wavelengths, progressively tracing warmer
dust close to the star.

\begin{figure} 
\includegraphics[width=10cm]{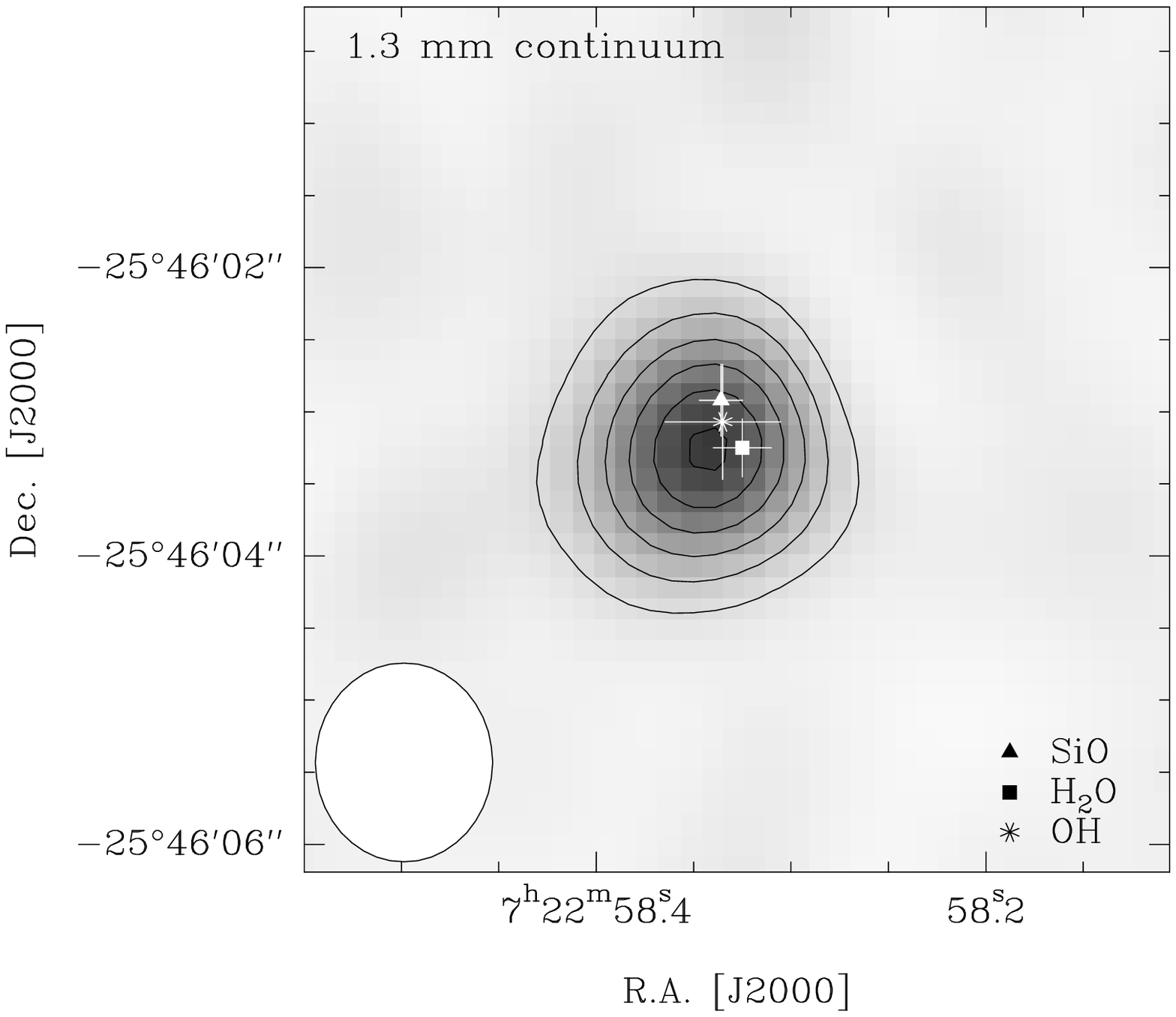}
\caption{1.3 mm continuum emission map of VY~CMa. The synthesized beam is 1.38$\arcsec$ x
1.23$\arcsec$ (PA = 169$\degr$) as indicated by the ellipse in the bottom left corner.
Contour levels are every 40 mJy/beam (5$\sigma$). 
The positions of the SiO \citep{wri90}, H$_2$O \citep{bow93} and OH \citep{bow83}
maser centroids are indicated.
The 1.3 mm continuum peak is located at R.A. = 07$^{\rm h}$22$^{\rm m}$58$\fs$339 and
Dec. = $-$25$\degr$46$\arcmin$03$\farcs$24 (J2000).}
\label{figcont}
\end{figure}

As seen from Fig.\ref{figcont}, the continuum emission is not resolved by our synthesized beam
($\sim$ 1.4$\arcsec$). The continuum visibilities are well fitted by a point source of 288 $\pm$
25 mJy (with 3$\sigma$ rms), located at a position +0.15$\arcsec$ $\pm$ 0.01$\arcsec$ to the east
and $-$0.05$\arcsec$ $\pm$ 0.01$\arcsec$ to the south, relatively to the phase center. Our
measurement of the flux density is consistent with the value of 270 $\pm$ 40 mJy from previous
SMA observations at 215 GHz \citep{shi04} but differs from the previous bolometric values of
630 $\pm$ 65 mJy obtained with the JCMT \citep{kna93} and 378 $\pm$ 20 mJy with the SEST
\citep{wal91}. Part of the difference could be explained by contamination from line emission
in the bolometric observations. We note, however, that even the $^{12}$CO(2-1) line, presumably
the strongest in this frequency band, contributes only for about 2\% and 7\% of the total flux
density measurements with the JCMT and SEST, respectively. On the other hand, we cannot exclude
the presence of an extended low surface brigthness emission that could have been missed in our
observations. Independant checks of the continuum flux density in the upper and lower side bands
give consistent values within 10\%, implying that our measurements are not severly contaminated
by weak line emission.

Because the continuum emission is of order of $\sim$ 1 mJy in the centimeter wavelenghts
\citep{lip05} and given that the contribution of the central star (black body emission)
amounts for S$_*$ $\sim$ 10 mJy, {\em i.e.} is negligible with respect to the total 1.3 mm
continuum emission S$_\nu$, this latter should arise almost completely from dust thermal emission.
Under the Rayleigh-Jeans approximation and if the dust emission is optically thin, we can
estimate the dust mass M$_{d}$ within our 1.4$\arcsec$ beam, according to:
\begin{equation}
M_{d} = \frac{2 c^2 D^2 a_g \rho_g (S_\nu - S_*)}{3 Q_\nu k T_d \nu^2}
\end{equation}
\noindent where D is the distance to the source, $\rho_g$ and a$_g$ are the mass density and
radius of the dust grains, Q$_\nu$ is the grain emissivity and $T_d$ the average temperature
of the dust envelope. We assume typical values for oxygen rich stars, a$_g$ = 0.2 $\micron$,
$\rho_g$ = 3.5 g.cm$^{-3}$, Q$_\nu$ = 5.65 10$^{-4}$($\nu$ / 274.6 GHz), following 
\citet{kna93}. The ratio of IR fluxes indicates a temperature T$_d$ $\sim$ 250 K in the dust
shell \citep{lip05}. Accordingly, we derive a dust mass of 1.5 10$^{-3}$ M$_\odot$, close
to the value M$_d$ = 1.3 10$^{-3}$ M$_\odot$ given by \citet{lip05}.
Assuming that the dust is uniformly distributed within a radius comparable to
our synthesized beam (upper limit of 1.4$\arcsec$), the optical depth towards
the central star would be $\tau_{1.3mm}$ $\sim$ 0.01.

Furthermore, assuming an expanding velocity V$_{\rm exp}$ = 15 km~s$^{-1}$ (similar to the
escape velocity of $\sim$ 20 km~s$^{-1}$ at the dust formation radius, see also \S\ref{model})
leads to an age of order of 300 years. The average mass loss rate over this timescale is
4.7 10$^{-4}$ M$_\odot$/yr, assuming an isotropic mass loss and a dust-to-gas ratio of 100.
This value is slightly higher than previous measurements for VY~CMa (see {\em e.g.}
\citealt{dan94,har01}) and comparable to mass loss rates estimated for other supergiants
\citep{net87}. \citet{smi01} and \citet{hum05} showed that localized and short-term
ejection events in VY~CMa could in fact reach 10$^{-3}$ M$_\odot$/yr.

On the other hand, as pointed out by the referee, there is a possibility that the continuum
dust emission originates from a circumstellar disk, as advocated earlier by \citet{bow83} and
\citet{ric98} from observations of OH and H$_2$O maser emissions. However, the angular resolution
of our current data is not enough to constrain this possibility.

\subsection{Molecular line emission}

\subsubsection{Description of the molecular line emission} \label{molecularline}

\begin{figure} 
\includegraphics[height=11cm]{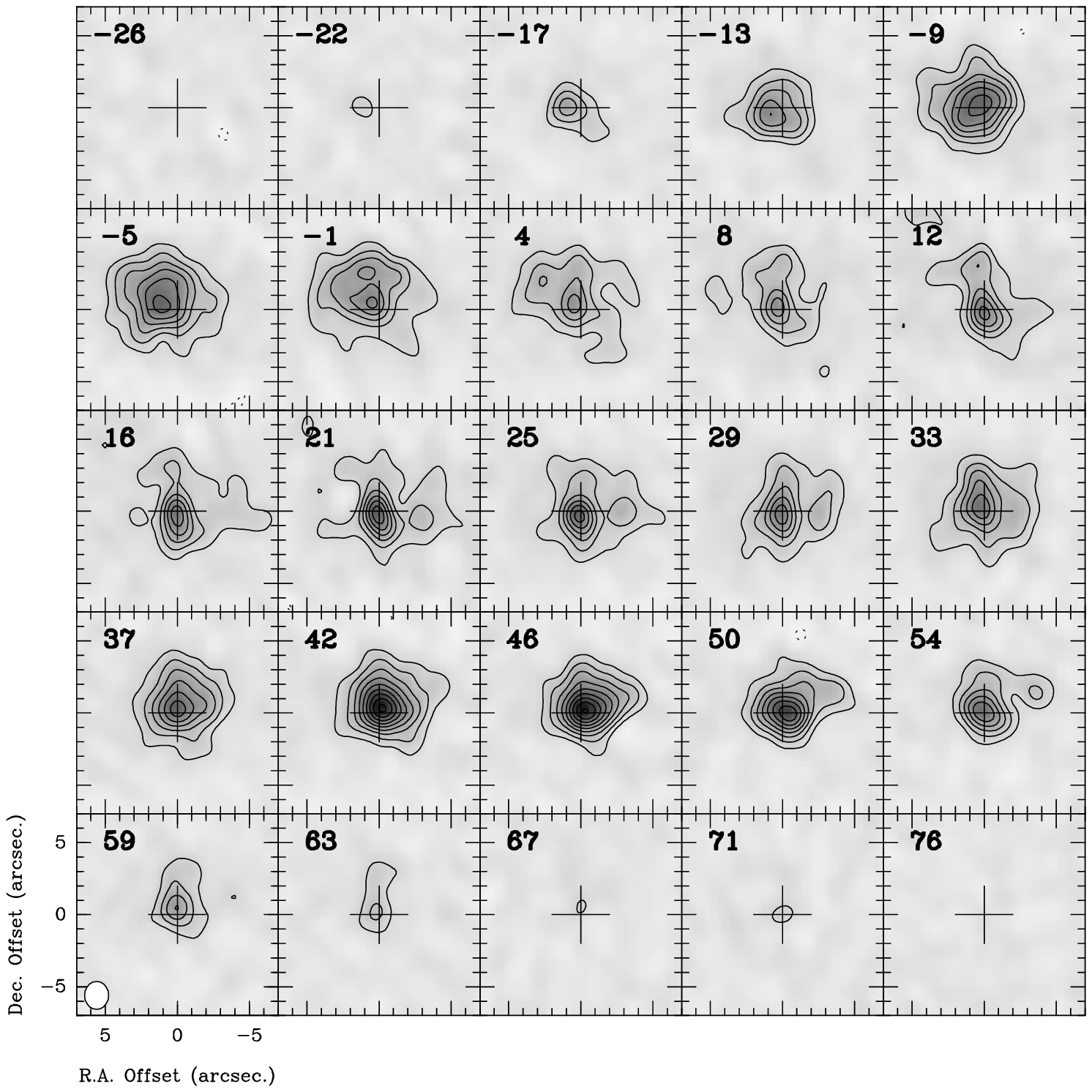}
\caption{Channel maps of the $^{12}$CO(2-1) line emission. The synthesized beam sizes
1.94$\arcsec$ x 1.67$\arcsec$ (PA = 0$\degr$) and is shown in the bottom left corner of
the figure. Contour levels are drawn every 0.55 Jy/beam or 3.9 K ($\sim$ 4$\sigma$).
The velocity resolution is 4.2 km~s$^{-1}$. The cross indicates the position of
the phase center.} \label{fig12co}
\end{figure}

The $^{12}$CO(2-1) emission (Fig.\ref{fig12co}) extends to $\sim$ 7$\arcsec$ around the central
star and is detected over a (LSR) velocity range $-$25 to 70 km~s$^{-1}$. There are three
distinct emission peaks in the channel maps: a first one at a velocity of about $-$5 km~s$^{-1}$
with an offset of 1$\arcsec$ to the east, a second one close to the systemic velocity of VY~CMa
(V$_{SYS}$ $\sim$ 22 km~s$^{-1}$), centered on the stellar position and a third one around +45
km~s$^{-1}$ with an offset of 1$\arcsec$ to the west. The line emission is stronger in the
western redshifted part of the nebula, where it reaches $\sim$ 5 Jy/beam. The velocity channels
around the systemic velocity of VY~CMa (12 $<$ V$_{LSR}$ $<$ 33 km~s$^{-1}$) show a north-south
elongated structure of roughly 5$\arcsec$ x 3$\arcsec$, embedded in a more diffuse envelope.
A second weak peak is seen 3$\arcsec$ to the west.

\begin{figure} 
\includegraphics[height=8cm]{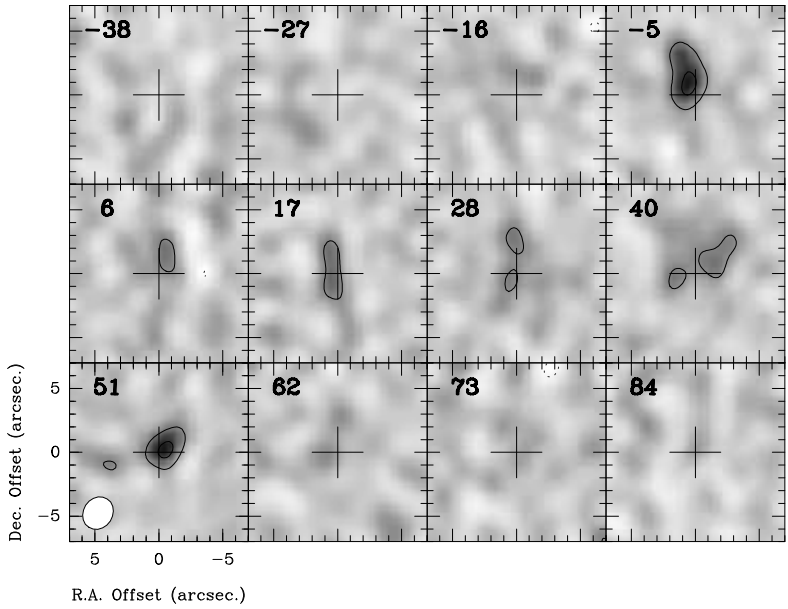}
\caption{Channel maps of the $^{13}$CO(2-1) line emission. The synthesized beam sizes
2.65$\arcsec$ x 2.30$\arcsec$ (PA = 147$\degr$) and is shown in the bottom left corner
of the figure. Contour levels are drawn every 0.2 Jy/beam or 0.82 K ($\sim$ 3$\sigma$).
The velocity resolution is 11 km~s$^{-1}$. The cross indicates the position of the phase
center.} \label{fig13co}
\end{figure}

\begin{figure} 
\includegraphics[height=11cm]{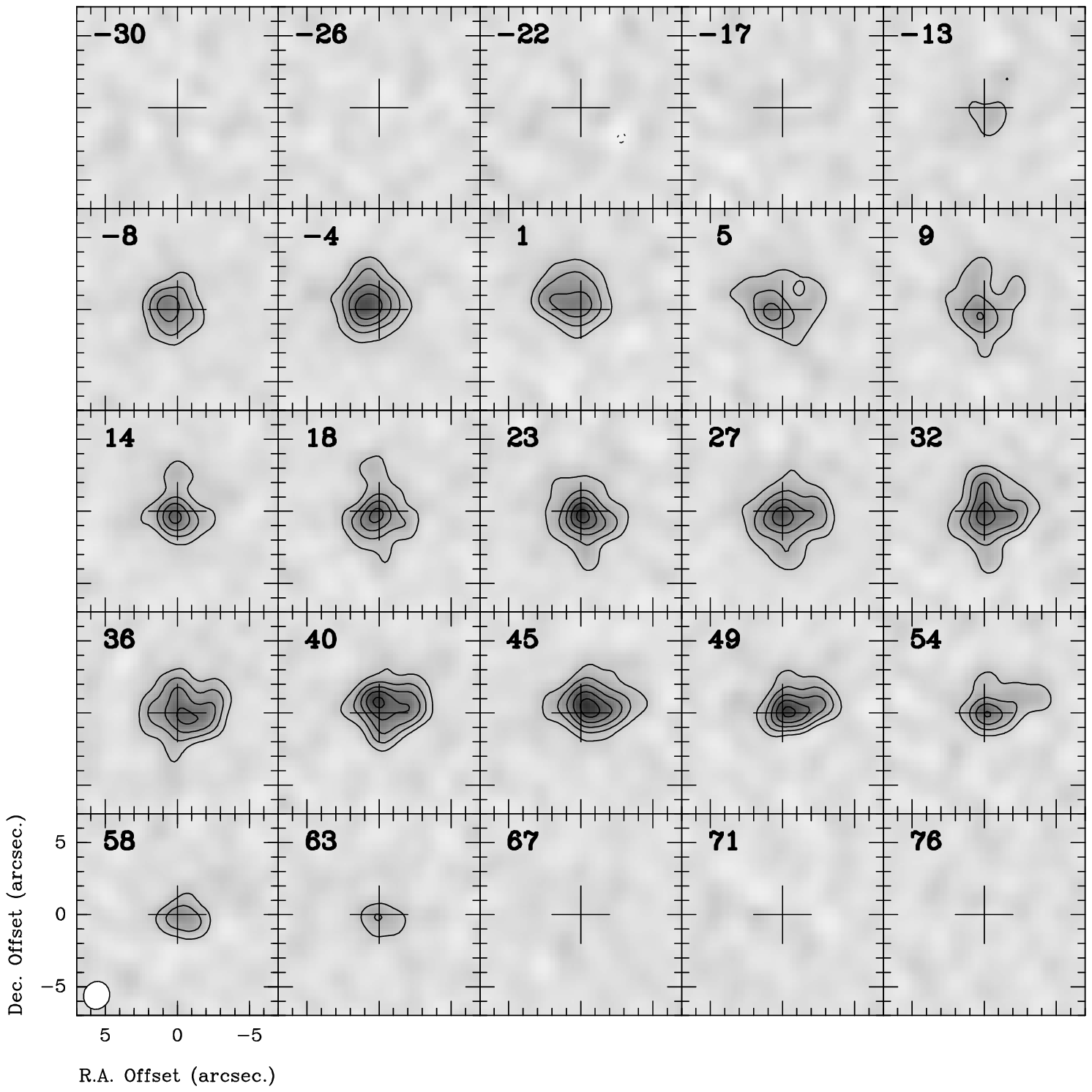}
\caption{Channel maps of the SO(6$_5$-5$_4$) line emission. The synthesized beam sizes
1.99$\arcsec$ x 1.80$\arcsec$ (PA = 155$\degr$) and is shown in the bottom left corner
of the figure. Contour levels are drawn every 0.44 Jy/beam or 3.1 K ($\sim$ 4$\sigma$).
The velocity resolution is 4.4 km~s$^{-1}$. The cross indicates the position of the phase
center.} \label{figso}
\end{figure}

\begin{figure} 
\includegraphics[width=15cm]{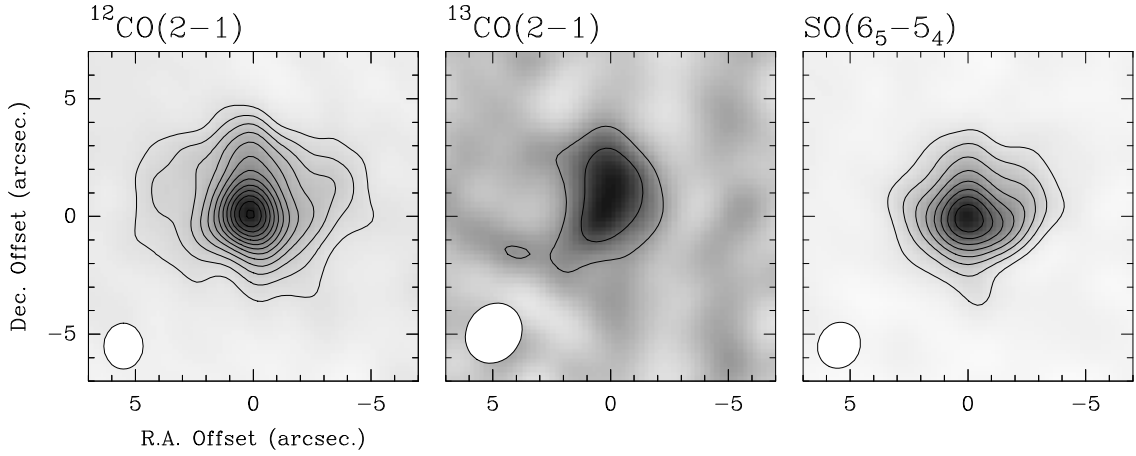}
\caption{Integrated maps of the $^{12}$CO(2-1), $^{13}$CO(2-1), SO(6$_5$-5$_4$) emission
lines. The synthesized beam sizes are indicated in the bottom left corner of each frame.
Contour levels are every 140 (7$\sigma$), 30 (3$\sigma$), and 105 (7$\sigma$) K~km~s$^{-1}$
respectively. Offsets refer to the position of the phase center.} \label{figint}
\end{figure}

Our $^{13}$CO(2-1) channel maps (Fig.\ref{fig13co}), although of low signal-to-noise, present
globally the same features as seen in $^{12}$CO(2-1). The two peaks are well detected at
velocities $\sim$ $-$5 and +50 km~s$^{-1}$, and separated by $\sim$ 2$\arcsec$. The emission
around the systemic velocity is weakly detected. The channel maps of the SO(6$_5$-5$_4$) emission
(Fig.\ref{figso}) also show the same features as that of the $^{12}$CO(2-1). The integrated
intensity maps for the three species are presented in Fig.\ref{figint}.

The morphology and kinematics of the $^{12}$CO(2-1) emission around VY~CMa do not reproduce
those of a typical spherical expanding shell. We interpret the east and west offset bright
peaks, with high velocity relatively to the systemic velocity, as the signature of a east-west
bipolar outflow. \citet{shi03} already suggested the presence of a bipolar outflow around VY~CMa
from the butterfly like shape of the SiO(1-0) emission. The idea was further supported by SiO
masers linear polarization measurements \citep{shi04}, which result in a mean polarization
position angle of 72$\degr$, in the axis of the SiO bipolar structure. Previous observations of
OH \citep{bow83,bow93} and H$_2$O \citep{ric98} masers also show some elongation in a
similar direction with position angles of 50$\degr$ and 60$\degr$, respectively. However, the
maser emission is clumpy and might not trace the complete geometry of the outflow.

\begin{figure} 
\includegraphics[height=7cm]{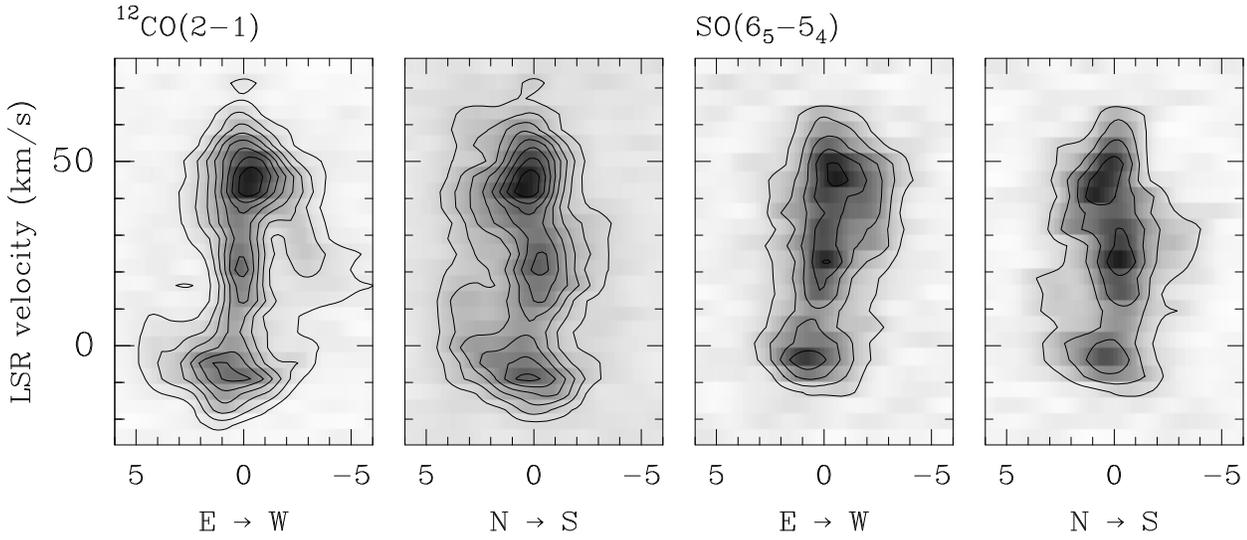}
\caption{Position - velocity diagrams along the east-west and north-south directions for
the $^{12}$CO(2-1) and SO(6$_5$-5$_4$) line emissions. Countour levels every 0.60 and 0.48
Jy/beam (4$\sigma$), respectively.}
\label{figpv}
\end{figure}

In addition, the 1.3 mm continuum and CO emission around the systemic velocity indicate the
presence of a central dense component. The position-velocity diagrams (Fig.\ref{figpv}), along
the east-west and north-south directions, {\em i.e.} along and perpendicular to the outflow axis,
support our interpretation. Along the east-west direction, the high velocity bipolar outflow and
the slowly expanding shell appear clearly as two distinct kinematic components. The bipolar
outflow emission shows a point symmetry relative to the stellar position. It extends to roughly
5$\arcsec$ from the star and has a velocity up to 45 km~s$^{-1}$ with respect to the systemic
velocity. This corresponds to a timescale of about 800 yr, comparable to the past period of strong
mass loss activity \citep{hum05,dec06}. The slowly expanding envelope can be seen out to a
radius of roughly 5$\arcsec$. Given an expansion velocity of 15 km~s$^{-1}$ (see \S\ref{model}),
the corresponding timescale is 2400 yr. The overall features are similar in the $^{12}$CO(2-1) and
SO(6$_5$-5$_4$) position-velocity diagrams, indicating that they arise from the same regions.

\subsubsection{Comparison with single dish spectra} \label{molecularspectra}

\begin{figure} 
\includegraphics[width=10cm]{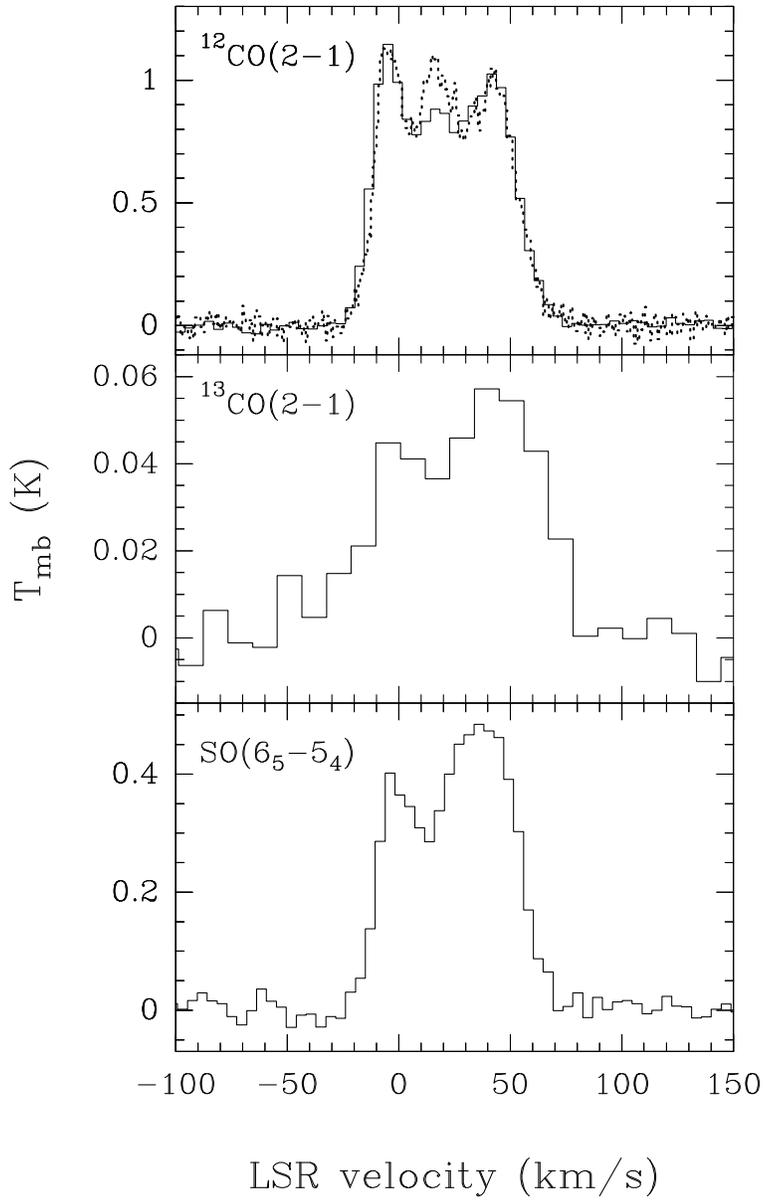}
\caption{Spectra of the $^{12}$CO(2-1), $^{13}$CO(2-1) and SO(6$_5$-5$_4$) lines smoothed to a
19$\farcs$7 FWHM beam. The $^{12}$CO(2-1) spectrum observed with the JCMT by \citet{kem03} is
shown as a dashed line in the upper panel. The velocity resolution is 4.2, 11 and 4.4 km~s$^{-1}$
from top to bottom.}
\label{figspec}
\end{figure}

We present in Fig.\ref{figspec} the spectra of the $^{12}$CO(2-1), $^{13}$CO(2-1) and
SO(6$_5$-5$_4$) lines, taken at the central position and convolved to a beam size of FWHM
19$\farcs$7, equal to the JCMT beam at 230 GHz. The $^{12}$CO(2-1) line profile obtained from our
interferometric data can then be directly compared to the single-dish $^{12}$CO(2-1) spectrum
\citep{kem03} at the same angular resolution. The $^{12}$CO(2-1) SMA spectrum shows three
peaks at V$_{LSR}$ $\sim$ $-$5, 20 and 45 km~s$^{-1}$, similar with the JCMT spectrum. The high
velocity channels (V$_{LSR}$ $<$ 5 and $>$ 35 km~s$^{-1}$) reach the same brightness temperature
as the single-dish observations, indicating that the interferometer recovered all of the emission
for these velocities. Close to the systemic velocity of VY~CMa, {\em i.e.} for 5 $<$ V$_{LSR}$ $<$ 30
km~s$^{-1}$, there is a small difference between the two profiles, although less than $\sim$ 20\%
of the emission is lost by the interferometer. We estimate the integrated intensity of the
$^{12}$CO(2-1) line to 65 K~km~s$^{-1}$, comparable to the value of 66 $\pm$ 7 K~km~s$^{-1}$
reported by \citet{kem03}. The integrated intensities of the $^{13}$CO(2-1) and SO(6$_5$-5$_4$)
lines are estimated to 4 K~km~s$^{-1}$ and 28 K~km~s$^{-1}$, respectively. The 3$\sigma$ rms
uncertainty is $\sim$ 1 K~km~s$^{-1}$.

\subsubsection{SO emission}

While carbon chemistry is mostly absent in the circumstellar envelope (CSE) of O-rich stars,
most of the carbon being locked up in CO, sulfur chemistry is expected to be relatively active.
S-bearing species are indeed commonly observed in the massive CSE around O-rich stars
\citep{sah92, omo93}. 

Under chemical thermodynamical equilibrium, \citet{tsu73} showed that the sulfur in CSEs is mostly 
in the form of H$_2$S. Later non-equilibrium chemical models \citep{sca80, nej88, wil97}
predict that a large proportion of sulfur should go into SO and SO$_2$, as a result of the reactions:
S + OH $\longrightarrow$ SO + H, HS + O $\longrightarrow$ SO + H and SO + OH $\longrightarrow$
SO$_2$ + H. OH is a natural product of the photodissociation of H$_2$O, and VY~CMa exhibits intense
maser emission for these two species \citep{bow83,ric98}. On the other hand, the production of
HS requires highly endothermic reactions. From our SMA observations, the SO and CO morphology and
kinematics are similar, implying that SO is present both close to the star and in the bipolar
outflow. Observations of \citet{san00} towards the bipolar nebula OH~231.8+4.2 show a similar trend.
These authors argue for the presence of SO both in the central component and in the accelerated lobes, 
with no significant difference in abundance between the two components.

Assuming optically thin emission and an excitation temperature of T$_{ex}$ = 40 K \citep{sah92},
we estimate a SO column density $\sim$ 10$^{16}$ cm$^{-2}$ towards the center. The
corresponding SO abundance relative to H$_2$ is $\sim$ 10$^{-6}$.

High angular resolution observations of other S-bearing species such as H$_2$S and SO$_2$ would
provide a better understanding of the sulfur chemistry and put stronger constraints for chemical
models.

\section{A simple model of the envelope} \label{model}

We have constructed a simple model in order to better understand the geometry and physical
conditions within the circumstellar envelope around VY~CMa. The combination of our high angular
resolution CO(2-1) data together with previous single dish observations of multiple CO rotational
transitions \citep{kem03} provides very useful constraints for our model and leads us to a
coherent picture of the envelope.

\begin{figure} 
\includegraphics[width=13cm]{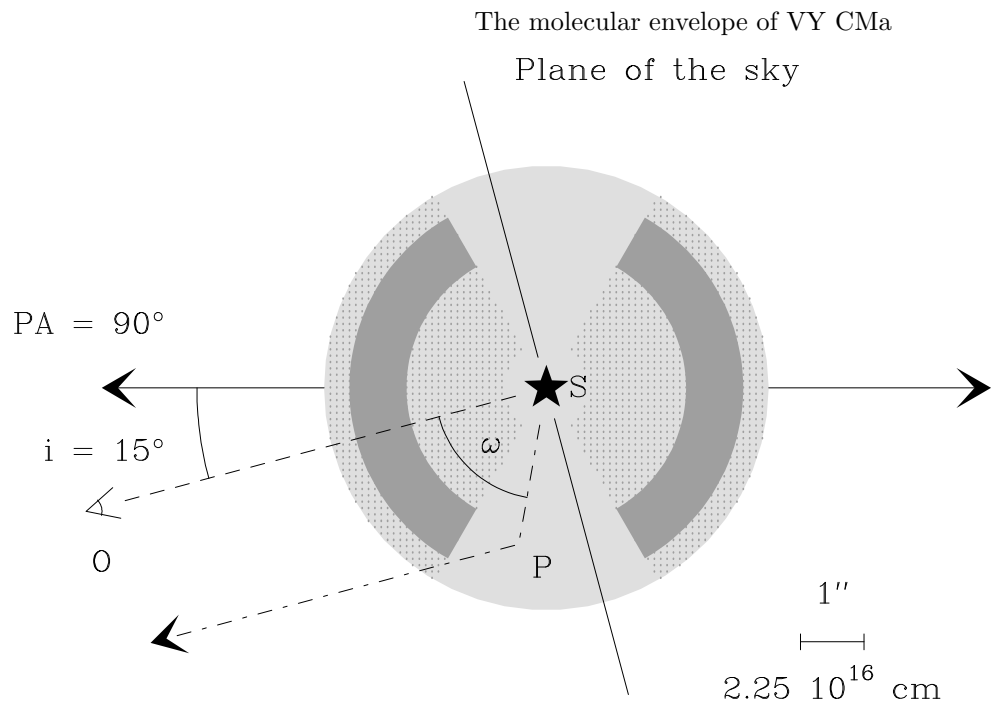}
\caption{Sketch of the geometry of our model. The central star VY~CMa (S) is embedded in a spherically
symmetric slowly expanding envelope (light grey). Part of the nebula is occupied by a bipolar outflow
with a wide opening angle (dotted pattern). Regions of density enhancement within the bipolar lobes
are also included (filled bands). The line of sight OS intercepts the bipolar outflow axis with an
inclination angle of 15$\degr$. A stellar photon is scattered at point P inside the envelope to the
observer. The angle $\omega$ is defined as $\hat{OSP}$.}
\label{figsketch}
\end{figure}

The geometry of our model is illustrated in Fig.\ref{figsketch}. The envelope is assumed to be
axisymmetric and radially expanding. The bipolar outflow direction defines the symmetry axis
of the envelope. We project the envelope directly into a regular 3-D grid. At each grid point,
the gas density, kinetic temperature and expansion velocity can be specified. The emerging CO
intensity along each ray through the envelope is calculated by integrating directly the standard
radiation transfer equation. We take into account 15 rotational levels of the CO molecule in its
vibrational ground state. The populations on the different rotational levels, which are necessary
for the calculation of the line opacity and source function at each grid point, are determined by
solving the statistical equilibrium equations within the framework of the large velocity gradient
formalism. The collision rates between CO and molecular hydrogren are taken from \citet{flo85} and
calculated for different temperatures using the prescription of \citet{jon75}. Because the envelope
around VY~CMa is oxygen rich, we adopt an abundance f = [CO]/[H$_2$] = 3 10$^{-4}$ as used in the
analysis of \citet{kem03}. The local line profile, which is needed to integrate the radiative
transfer equation, is determined through the turbulence velocity of the molecular gas. We assume
a turbulence velocity of 1 km~s$^{-1}$ in the slowly expanding envelope and 3 km~s$^{-1}$ in the
bipolar lobes, respectively. This assumption is reasonable as a higher outflow velocity in the
lobe would result in more turbulent environment.

As discussed in \S\ref{molecularline}, two distinct kinematic components are clearly present in the
nebula around VY~CMa: (1) a slowly expanding envelope slightly elongated in the north-south direction
surrounded by a diffuse halo with some extension toward the north-west direction and (2) a high
velocity bipolar outflow oriented in the east-west direction. However, to keep our model as simple
as possible, we assume that the slowly expanding component is spherically symmetric. We adopt an
expansion velocity of 15 km~s$^{-1}$ for this component, which allows us to satisfactorily fit the
central peak seen in single-dish CO observations. The presence of diffuse emission suggests
that the gas density in the slowly expanding envelope peaks more sharply toward the center than for
the case of constant mass loss. 
We therefore adopt a simple density distribution n($r$) = 4.5 10$^5$ (10$^{16}$ cm/$r$)$^{2.5}$, which
deviates slightly from the usual density distribution generated by constant mass loss. This adopted
distribution corresponds to a mass loss rate $\dot{M}$ = 4.5 10$^{-5}$ (10$^{16}$ cm/$r$)$^{0.5}$
M$_\odot$/yr, or a decrease in mass loss rate by a factor of 4 between the inner radius (R$_{\rm in}$
and the outer radius (R$_{\rm out}$). The kinetic temperature of the molecular gas in the oxygen-rich
circumstellar envelopes with constant mass loss has been known to roughly follow a power law T($r$)
$\sim$ $r^{-\alpha}$ from the modelling works of \citet{gol76} and \citet{jus94}. Although the mass
loss rate in the slowly expanding envelope of VY~CMa is not strictly constant, resulting probably in
some slight change in the temperature profile in comparison to the simple constant mass loss case,
we expect that qualitatively, the temperature profile still follows the power law. Therefore, we
assume in our calculations, for the sake of simplicity, that the kinetic temperature varies as
T($r$)= 500 K (2.25 10$^{15}$ cm/$r$)$^{0.7}$, which is similar to that used by \citet{kem03} 
to model the oxygen-rich circumstellar envelopes.

\begin{figure} 
\includegraphics[width=10cm]{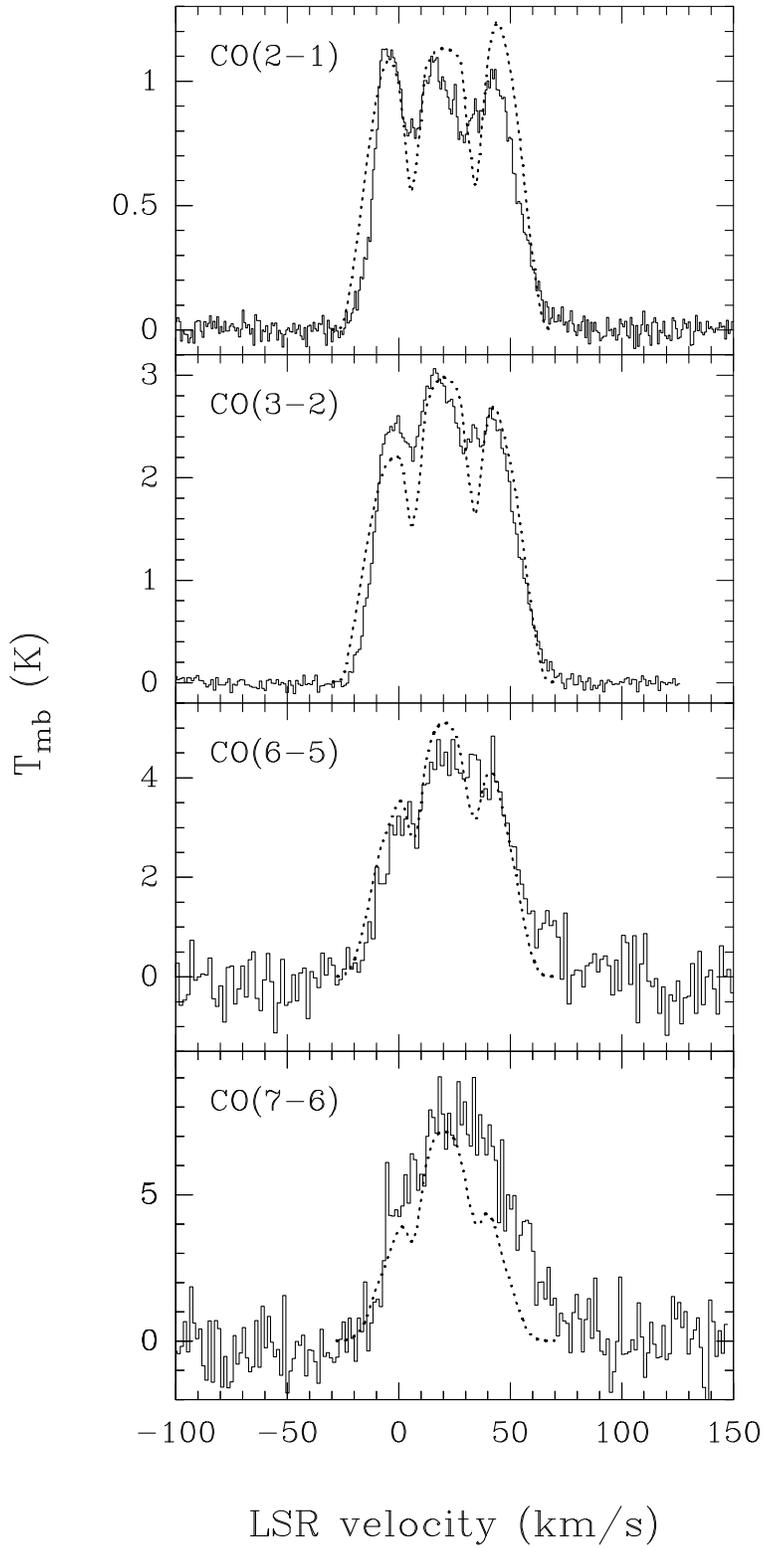}
\caption{Spectra of the CO(2-1), CO(3-2), CO(6-5) and CO(7-6) lines observed by \citet{kem03}
using JCMT are shown (full lines) together with the predicted CO line profiles from our model
(dotted lines).}
\label{figspecco}
\end{figure}

In order to reproduce the velocity gradient seen in the position-velocity
diagrams (Fig.\ref{figpv}), the outflow velocity in the bipolar lobes is allowed to increase linearly
with radial distance from the central star. In addition, the outflow velocity in the bipolar lobes
also varies as a function of the polar angle $\theta$ measured from the bipolar axis. The strong
emission peaks at large velocity shifts suggest the presence of an enhanced density shell within the
bipolar lobes. However, this density enhancement is more pronounced in the receeding lobe as clearly
seen in the position-velocity diagram. With the density enhancement included in the model, we find
that the bipolar lobes and the slowly expanding shell can account for the highly unusual triple peak
profile of single-dish CO lines observed by \citet{kem03} (see Fig.\ref{figspecco}).

The intensities of the CO lines calculated along individual lines of sight are used to form the
model channel maps (Fig.\ref{figchanmodel}). We convolve these model channel maps with a gaussian
beam with specified FWHM to produce the model CO line profiles, which are indicated by the dashed
lines and compared with the JCMT observations of \citet{kem03} in Fig.\ref{figspecco}.

We infer from our modelling that the velocity of the gas in the lobes increases linearly from 
15 km~s$^{-1}$, which is similar to the expansion velocity of the spherical envelope, to
45 km~s$^{-1}$ at the outer radius. The bipolar outflow is found to be very open, with an opening
angle of 120$\degr$, and inclines at 15$\degr$ from the line of sight. All the model parameters,
which provide a reasonable fit to both the SMA and single dish data, are shown in Table 3.

Although relatively simple, our model provides a qualitative understanding of the single dish data
obtained by \citet{kem03} and also of the structure of the envelope (Figs.\ref{figspecco},
\ref{figchanmodel}, \ref{figpvmodel}). The high angular resolution SMA data show that the CO(2-1)
emission originating from the high velocity bipolar lobes produces a double horn profile at roughly
$\pm$20 km~s$^{-1}$ from the systemic velocity. The central broad peak of the CO(2-1) line comes
from the slow component. Higher transitions of CO probe progressively warmer and denser gas located
in the inner part of the envelope. However, both the density and temperature of the gas in the inner
region of the bipolar outflow, as suggested by our model, are lower than that in the slowly
expanding envelope. As a result, the central peak becomes more prominent in higher transitions while
the strength of the emission from the bipolar lobes decreases noticeably. In addition, the lower
outflow velocity in the inner part of the bipolar lobes also leads to a narrower full width at zero
velocity for higher transitions of CO. This behavior can be clearly seen in Fig.\ref{figspecco}.

\begin{figure} 
\includegraphics[height=7cm]{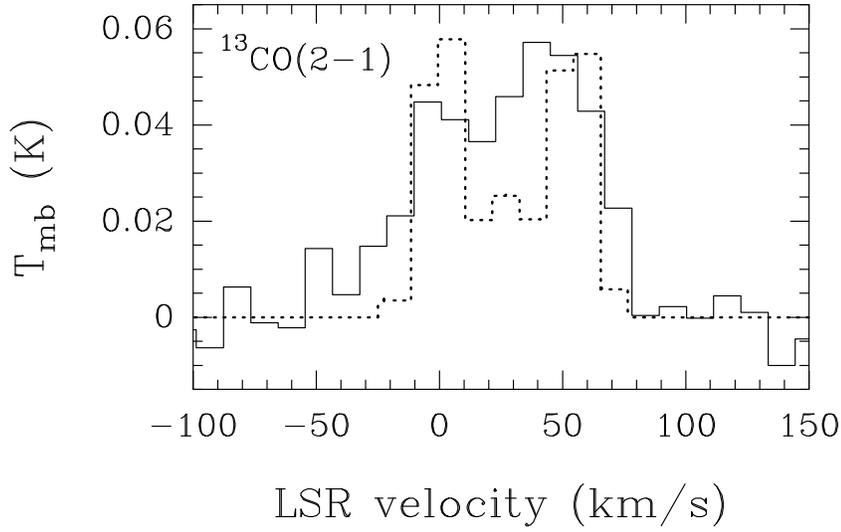}
\caption{SMA spectrum of the $^{13}$CO(2-1) line convolved with a 19$\farcs$7 FWHM beam (full
line) together with the output spectrum from our model (dotted line). The model spectrum is
smoothed to the same velocity resolution of the SMA data (11 km~s$^{-1}$). A ratio
$^{12}$C/$^{13}$C = 60 was inferred from our modelling.}
\label{fig13comodel}
\end{figure}

\begin{figure} 
\includegraphics[width=10cm]{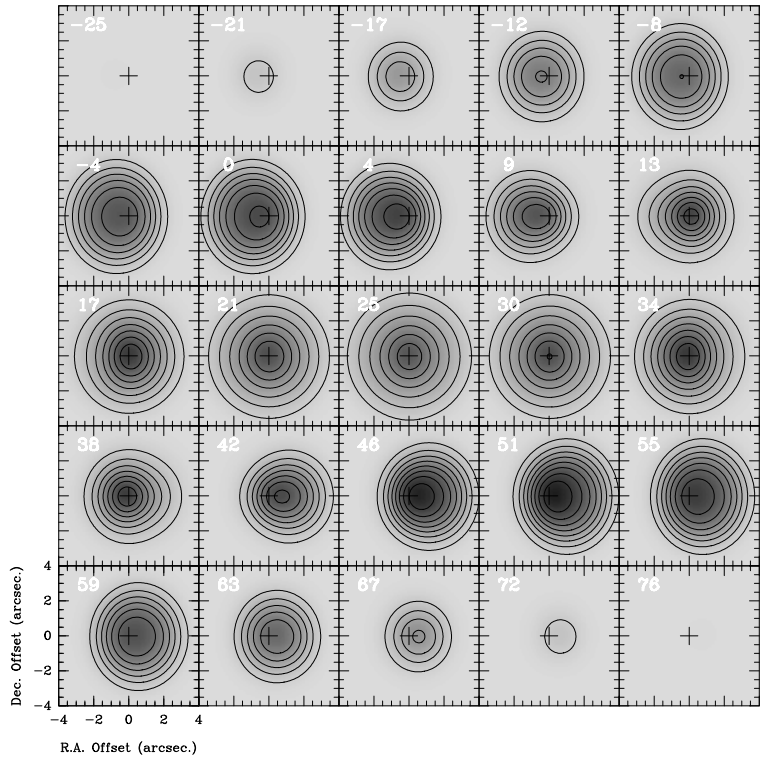}
\caption{Predicted channel maps of the CO(2-1) emission from the envelope around VY~CMa.
Contour levels are the same as in Fig.\ref{fig12co}.}
\label{figchanmodel}
\end{figure}

\begin{figure} 
\includegraphics[height=7cm]{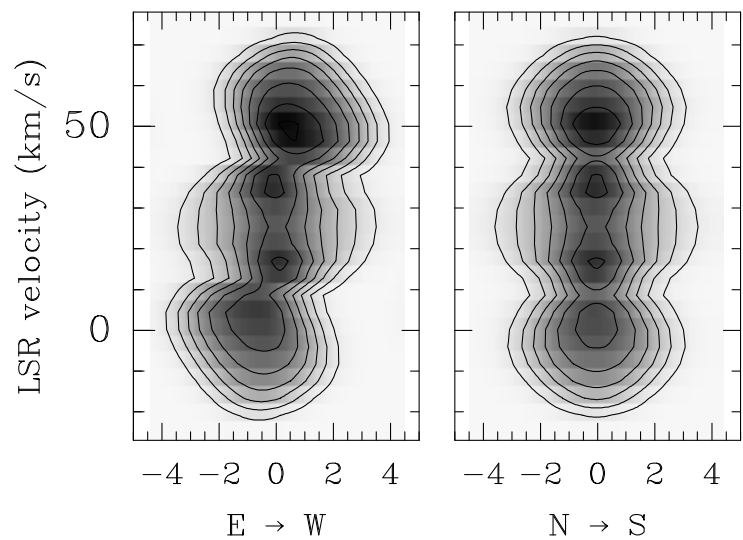}
\caption{Predicted position-velocity diagram of the CO(2-1) emission along the east-west (left frame)
and north-south (right frame) direction.}
\label{figpvmodel}
\end{figure}

By comparing the model prediction for the strength of the $^{13}$CO(2-1) line with our SMA
observations, we estimate the isotopic ratio $^{12}$C/$^{13}$C = 60.
This ratio is significantly larger than the values $^{12}$C/$^{13}$C = 6 and 12
found in the red supergiants $\alpha$ Ori and $\alpha$ Sco, respectively \citep{har84, hin76}.
It is likely that, prior to the current mass loss episode, VY~CMa experienced no significant
mixing of H burning products to the surface.
Our model predicts that most of the $^{13}$CO(2-1) line emission should arise from the bipolar
outflow (Fig.\ref{fig13comodel}), which is consistent with our SMA imaging data (Fig.\ref{fig13co}).

We emphasize here that our model is constructed to explain the general features of the high
resolution SMA data and single-dish CO observations. Other features such as the slight elongation
of the slowly expanding shell and the more complicated structure within the bipolar lobes are not
considered. However, our model could serve as a useful starting point when higher angular
resolution data of CO(2-1) and higher transitions become available in the future.

\section{Discussion}

\subsection{Connection with optical imaging/spectroscopy data}

Our high angular resolution CO(2-1) data bring a clearer overall picture of the molecular
envelope around the enigmatic red supergiant VY~CMa. The envelope consists of two kinematic
components: a slowly expanding shell slightly elongated in the north-south direction and a high
velocity bipolar outflow oriented in the east-west direction with small inclination angle to the
line of sight. The presence of these two kinematic components have already been hinted in previous
observations of H$_2$O and OH masers \citep{bow83, bow93, ric98}. Given the small inclination
angle $\sim$ 15$\degr$ of the bipolar outflows to the line of sight, as derived from our model and
assuming uniform illumination in the nebula, one would expect that the envelope appears only
slightly asymmetric with a brighter east lobe (nearer side). However, optical and near IR images
obtained by \citet{mon99} and \citet{smi01} all show a highly irregular and asymmetric nebula around
VY~CMa. The nebula is seen to the west and southwest of the bright point-like source identified as
the obscured central star VY~CMa and is likely the result of a non-isotropic illumination by the
central star. Comparing the optical images with our SMA data, we note that the optical nebulosity
corresponds to the location of the west (far side) lobe and just a small part of the east (near side)
lobe. It is therefore likely that the light from VY~CMa escapes preferentially in the direction of
the west bipolar lobe and scattering on dust brings a fraction of the light back toward the observer.

The strong 1.3 mm continuum emission detected in our data (see \S\ref{cont}) confirms the presence
of a thick inner dust shell surrounding VY~CMa. The average optical depth would be so high, if the
distribution of material in the dust shell is smooth, that all the stellar light would be completely
extincted. The complex nebula seen in the optical and near IR then suggests that stellar light must
be able to escape through some opening or holes in the dust shell, creating search light beams that
illuminate the nebula in an irregular pattern. We note that a similar suggestion was made by
\citet{wal78} to explain the subtle but peculiar change in the relative position between the star
and the brightest knot in the nebula.

The thick and compact inner dust shell detected in 1.3 mm continuum emission might be the region
where TiO and ScO bands \citep{her74, wal86} seen in emission originate. The calculated
excitation temperatures of these molecular bands range from 600 to 800 K, suggesting that they
come from a hot and dense region located close to the star.

High resolution optical and near IR images of \citet{mon99} and \citet{smi01} show a very
prominent long curved tail or arc to the north-west of VY~CMa together with several fainter
arcs and condensations in the envelope. The location of the long northwest arc roughly
coincides with the extended and somewhat diffuse CO(2-1) emission in the western part of
the envelope (Fig.\ref{fig12co}). Together with strong central emission core, this extended
emission is conspicuous in all channel maps redshifted with respect to the central star, most
prominent in the velocity range 21 -- 33 km~s$^{-1}$ and 46 -- 54 km~s$^{-1}$. The orientation
of this feature changes from west to north-west with increasing velocity upon closer inspection.
We can identify a similar feature in the channel maps of the SO emission at the same velocity,
although there are some slight differences at higher redshifted velocities, which are probably
related to local excitation of the SO molecule. It is therefore possible that the scattering of
stellar light by dust in this structure produces the prominent northwest arc seen in optical and
near IR images. Future observations of molecular lines at higher angular resolution and detailed
modelling of the light scattering by dust inside the envelope will help to clarify the relation
between the material distribution and the optical and near IR appearance of the nebula.
 
Recently, \citet{hum05} have obtained high resolution long slit spectroscopic data of VY~CMa.
By measuring the velocity of the reflected stellar absorption lines and K I emission lines they
suggest that the prominent northwest arc is expanding at a velocity of $\sim$ 50 km~s$^{-1}$ on
the plane of the sky. More surprising is the existence of diffuse gas and dust more or less
stationary with respect to the central star, mostly along the slit positioned to the southeast
of VY~CMa (slit V in their Fig.1). Thus, components with very different kinematics seem to
co-exist in the nebula. To reconcile the optical spectroscopic data with our SMA results, we
note that for the simple case of single scattering, the light scattered by moving reflective
material (dust particles) within the envelope is Doppler shifted by
V$_{\rm exp}$(1 $-$ cos$\omega$), where V$_{\rm exp}$ is the expansion velocity and $\omega$ is
the angle measured from the line of sight to the radial vector of the scattering point (see
Fig.\ref{figsketch}). The velocity shift can be seen to be largest in the west lobe, which is
receeding along the line of sight, because $\omega$ $>$ 90$\degr$. If the association of the
extended feature in the west lobe as traced by CO emission and the northwest arc is correct,
we would expect the velocity shift induced by scattering of moving dust to vary from
15 km~s$^{-1}$ for material in the arc moving close to the plane of the sky, which is part of
the slowly expanding shell, to $\sim$ 50 km~s$^{-1}$ or higher for material within the bipolar
lobe. This simple estimate of the velocity shift is consistent with the measurement of
\citet{hum05}. For absorption lines seen in scattering through the east lobe, the velocity
shift is expected to be close to zero because the angle $\omega$ is small. That might explain
the presence of dust and gas almost stationary with respect to the star as suggested by
\citet{hum05}. However, similar behavior seen in resonant K I lines are still difficult to
understand.

\subsection{Mass loss history}

There are strong evidences that the mass loss rate in VY~CMa is time variable, such as the
prominent arc-like features seen in optical \citep{smi01} and the relative strength of the CO
rotational line profiles \citep{dec06}.

Our modelling results suggest that the mass loss is varying as $\sim$ 4.5 10$^{-5}$
(10$^{16}$ cm/$r$)$^{0.5}$ M$_\odot$/yr in the slowly expanding envelope (V$_{\rm exp}$ =
15 km~s$^{-1}$), {\em i.e.} that the mass loss from VY~CMa is increasing with time, reaching
$\sim$ 6.4 10$^{-5}$ M$_\odot$/yr in the inner part of the envelope (R$_{\rm in}$) or about
100 years ago. In the bipolar lobes, the density enhancement between 5 -- 7 10$^{16}$ cm
represents an episode of strong mass loss about 350 to 500 years ago, for the expansion
velocity of 45 km~s$^{-1}$. The nominal mass loss rate at 6 10$^{16}$ cm is 10$^{-3}$
M$_\odot$/yr. A similar episode of intense mass loss has also been inferred from modelling
results of \citet{dec06}, although their assumption of spherical symmetry is inadequate to
describe the geometry and kinematics of the nebula around VY~CMa.

The current mass loss from VY~CMa is better traced by dust continuum emission in the mid
infrared or millimeter and sub-millimeter wavelengths. We derived a mass loss rate of
4.7 10$^{-4}$ M$_\odot$/yr from 1.3 mm dust continuum emission. This value is very close
to that obtained by \citet{har01} from fitting infrared spectroscopic data. As already noted
by \citet{dec06}, the difference between mass loss rate derived from CO line profiles and
dust continuum suggests that the dust continuum emission originates from the inner dense and
hot region (a few tens of stellar radii from the central star), which is not well traced by
CO low rotational lines.

\subsection{Origin of the complex envelope around VY~CMa}

The origin of the bipolar outflow and the north-south elongated structure in VY~CMa is still
unclear at present. However, we note the similarity in morphology (bipolar outflows and/or
disk-like structures) between the envelope of VY~CMa and that around several AGB and post-AGB
stars such as $\pi^1$ Gru \citep{chi06}, V Hya \citep{hir04}, M~2-56 \citep{cas01}
and X Her \citep{nak05}. The current popular model proposed by \citet{mor87} to explain the
striking bipolar morphology and the presence of collimated fast outflow in AGB and post-AGB
envelopes involves a low mass binary companion. The companion can have strong effect on the
structure of the envelope. Its gravitational pull can focus the AGB wind toward the orbital
plane to create an equatorial density enhancement. In addition, the accretion of wind material
into the companion leads to the formation of an accretion disk, which might drive the collimated
high velocity outflow. Although the overall morphology of the envelope is similar to that around
other AGB and post-AGB stars, the high velocity wind from VY~CMa, as suggested by our data and
modelling results, is only weakly collimated with very wide opening angle. It is conceivable
that the interaction between an unseen companion and VY~CMa could produce equatorial density
enhancement, possibly identified as the elongated structure and induce the collimated fast wind
to precess. Subsequent interaction with the nebula would result in wide bipolar lobe. The
prominent curved nebulous tail seen in scattered light and the extended and curved structure
seen in our CO(2-1) data might actually trace the precessing outflow. To explain the prominent
nebulous tail, \citet{mon99} also suggest the existence of a rotating jet and estimate the
rotation period of a few thousands years. That time scale might be related to the orbital period
of an unseen companion around VY~CMa. However, high resolution images from \citet{smi01} suggest
that this structure is actually much more complicated than expected from the model of \citet{mon99}
and probably more consistent with a localized ejection event on the stellar surface. Our currrent
SMA data cannot distinguish between these possibilities. Future high angular resolution observations
and monitoring are necessary to determine the distribution of material and kinematics in order to
test the presence of a companion.

VY~CMa might have magnetic activity, resembling solar flares or eruptions, as suggested by
\citet{smi01} in order to explain the presence of numerous arcs and bright knots seen in
scattered light in the nebula. We speculate further that the bipolar flow with wide opening
angle and the equatorial density enhancement could be similar to what is observed in the
solar wind. The fast solar wind originates from coronal holes located near the poles where
magnetic field lines are open and the slower but denser wind comes from the equatorial region.
The geometry of the solar wind is closely coupled to the dipolar field of the Sun. In the case
of VY~CMa, the presence and strength of the magnetic field can only be inferred indirectly
from observations of SiO, H$_2$O and OH masers. Observations of Zeeman effect in SiO maser
lines suggest a strength of 0 -- 2 G \citep{her06} for the magnetic field around
VY~CMa. Using simple estimates, \cite{sok98} and \cite{sok02} suggest that a magnetic field
of about 1 G can significantly influence the outflow.
The observed strength of the magnetic field
in VY~CMa thus suggests that it might affect the dynamics of the wind and if the field is organised
on large scale, {\em i.e.} dipolar configuration, it could explain the structure of the envelope
as seen in our SMA data.

\section{Conclusion}

In this paper, we present high angular resolution observations of the circumstellar molecular
envelope of the red supergiant VY~CMa obtained with the SubMillimeter Array. Thanks to the
wide frequency coverage of the SMA, we were able to trace the envelope in the $^{12}$CO(2-1),
$^{13}$CO(2-1) and SO(6$_5$-5$_4$) lines and dust continuum emission. We detected a strong and
unresolved continuum emission at the center of the nebula. The derived dust mass is 1.5 10$^{-3}$
M$_\odot$, implying a lower limit of the recent mass loss of 4.7 10$^{-4}$ M$_\odot$/yr over
the last 300 years. The CO(2-1) emission is resolved into two distinct kinematic components: a
slowly (15 km~s$^{-1}$) expanding shell elongated in the north-south direction and a high
velocity (45 km~s$^{-1}$) component with a clear velocity gradient along the east-west direction.
Similar structures are also seen in the $^{13}$CO(2-1) and SO(6$_5$-5$_4$) emission. We interpret
the high velocity component as a bipolar outflow with a wide opening angle ($\sim$ 120$\degr$),
for which the symmetry axis is oriented close to the line of sight (i = 15$\degr$). Using a
simple model, we could explain the main features of the CO(2-1) emission in our SMA data as well
as the previous single-dish CO multi-line observations. Density enhancement in the outflows,
representing a previous episode of high mass loss, is inferred from our model. Our data
complement previous high resolution optical spectroscopy and optical/IR images and provide a
better understanding of the complicated morphology of the nebula around VY~CMa. Future high
resolution and high sensitivity observations of CO lines, including high-J transitions, will
allow us to study the structure of the envelope and its physical properties in more detail.

\acknowledgements

We are very grateful to F. Markwick-Kemper for providing us her JCMT CO spectra on VY~CMa.
This work is supported in part by Academia Sinica and by grants from the National Science Council
of Taiwan (NSC) to D.-V-T. (NSC 94-2112-M-001-008), J.L., and S.K. This research made use of NASA's
Astrophysics Data System, as well as the SIMBAD database operated at CDS, Strasbourg, France.


\begin{table}
\begin{tabular}{lll}
\multicolumn{3}{l}{Table 1: Basic data of VY~CMa} \\
\hline
\hline
R.A. (J2000) & 07$^{\rm h}$22$^{\rm m}$58$\fs$3315 & (1) \\ 
Dec. (J2000) & $-$25$\degr$46$\arcmin$03$\farcs$174 & (1) \\
Distance & 1.5 kpc & (2) \\
Angular scale & 1$\arcsec$ $\sim$ 1500 AU $\sim$ 2.25 10$^{16}$ cm & \\
Luminosity L$_*$ & $\sim$ 2 10$^5$ L$_\odot$ & (3) \\
Effective temperature T$_{eff,*}$ & 2800 K & (3) \\
Stellar radius R$_*$ & 1.2 10$^{14}$ cm & (3) \\ 
Stellar mass M$_*$ & $\ge$ 15 M$_\odot$ & \\
Mass loss rate $\dot{M}$ & $\sim$ 2 -- 4 10$^{-4}$ M$_\odot$/yr & (4) \\
\hline
\end{tabular}
\tablerefs{ (1) from the Hipparcos catalogue, \citet{per97}; (2) \citet{lad78}; (3) \citet{mon99}; (4) \citet{bow83}, \citet{dan94}, \citet{sta95}, \cite{har01}.}
\end{table}


\begin{table}
\begin{tabular}{lccccc}
\multicolumn{3}{l}{Table 2: SMA line observations towards VY~CMa.} \\
\hline
\hline
\multicolumn{1}{c}{Line} & Rest freq. & Synthesized beam & P.A. & Vel. res. & 1$\sigma$ rms \\
                         & (GHz)      & (arcsec)         & (deg.) & (km~s$^{-1}$) & (mJy/beam)    \\
\hline
$^{12}$CO(2-1)  & 230.538 & 1.94 x 1.67 &   0 & 4.2 & 140 \\
$^{13}$CO(2-1)  & 220.399 & 2.65 x 2.30 & 147 &  11 & 65 \\
SO(6$_5$-5$_4$) & 219.949 & 1.99 x 1.80 & 155 & 4.4 & 110 \\
\hline
\end{tabular} 
\end{table}


\begin{table}
{\small
\begin{tabular}{lll}
\multicolumn{3}{l}{Table 3: Parameters of our model} \\
\hline
\hline
Envelope inner radius (R$_{\rm in}$) & 5 10$^{15}$ cm & \\
Envelope outer radius (R$_{\rm out}$) & 8 10$^{16}$ cm & \\
Bipolar lobe inner radius (R$_{\rm lobe}$) & 1.5 10$^{16}$ cm & \\
Opening angle of the bipolar lobes & 120$^o$ & \\
Envelope expansion velocity & 15 km~s$^{-1}$ & \\
Expansion velocity in the bipolar lobes & \multicolumn{2}{l}{[30 - 10 tan$\theta$/tan(60$^o$)]
(r - R$_{\rm lobe}$)/(R$_{\rm out}$ - R$_{\rm lobe}$) + 15 km~s$^{-1}$ $^\dagger$} \\
Gas density in the envelope & 4.5 10$^5$(10$^{16}$ cm/r)$^{2.5}$ cm$^{-3}$ & \\
Gas temperature in the envelope & 500 (2.25 10$^{15}$ cm/r)$^{0.7}$ K & \\
Gas density in the bipolar lobes & 6 10$^4$ cm$^{-3}$ & for R$_{\rm lobe}$ $<$ r $<$ 5 10$^{16}$ cm \\ 
...                              & 10$^5$ cm$^{-3}$ & for 5 10$^{16}$ cm $<$ r $<$ 7 10$^{16}$ cm \\
...                              & 10$^5$ (7 10$^{16}$ cm/r)$^2$ (cm$^{-3}$) & for r $>$ 7 10$^{16}$ cm \\
Gas temperature in the bipolar lobes & 150 (1.5 10$^{16}$ cm/r)$^{0.7}$ K & for R$_{\rm lobe}$ $<$ r $<$ 5 10$^{16}$ cm \\
...                              & 35 K & for 5 10$^{16}$ cm $<$ r $<$ 7 10$^{16}$ cm \\
...                              & 35 (7 10$^{16}$ cm/r) K & for r $>$ 7 10$^{16}$ cm \\
Inclination angle & 15$^o$ & \\
Position angle of the nebula axis & 90$^o$ & \\
f=[CO]/[H$_2$] & 3 10$^{-4}$ & \\
$^{12}$C/$^{13}$C & 60 & \\
Total molecular gas mass in the envelope & 0.15 M$_\odot$ & \\
\hline
\end{tabular}
}
\end{table}


\end{document}